\documentstyle[preprint,eqsecnum,aps,prb]{revtex}

\begin{document}
\title{Laser ablation of Co:ZnO films deposited from Zn and Co metal targets on
(0001) Al$_2$O$_3$ substrates.}
\author{W. Prellier\thanks{%
prellier@ismra.fr}, A. Fouchet, B. Mercey, Ch. Simon and B. Raveau}
\address{Laboratoire CRISMAT, CNRS\ UMR 6508, ENSICAEN,\\
6 Bd du Mar\'{e}chal Juin, F-14050 Caen Cedex, FRANCE.}
\date{\today}
\maketitle

\begin{abstract}
We report on the synthesis of high-quality Co-doped ZnO thin films using the
pulsed laser deposition technique on (0001)-Al$_2$O$_3$ substrates performed
in an oxidizing atmosphere, using Zn and Co metallic targets. We firstly
optimized the growth of ZnO in order to obtain the less strained film.
Highly crystallized Co:ZnO thin films are obtained by an alternative
deposition from Zn and Co metal targets. This procedure allows an homogenous
repartition of the Co in the ZnO wurzite structure which is confirmed by the
linear dependance of the out-of-plane lattice parameter as a function of the
Co\ dopant. In the case of 5\% Co doped, the film exhibits ferromagnetism
with a Curie temperature close to the room temperature.
\end{abstract}

\newpage

Diluted Magnetic Semiconductors (DMS) of III-V or II-VI types have been
obtained by doping semiconductors with magnetic impurities (Mn for example)%
\cite{1,2}. These materials are very interesting due to their potential
applications for spintronics\cite{3}. However, the low Curie temperature ($%
T_C$) has limited their interest\cite{4}. Based on the theoretical works of
Dietl {\it et al.}\cite{5}, several groups \cite{5a} have studied the growth
of Co-doped ZnO films\cite{6,7,8,9} which is a good candidate having a high $%
T_C$ \cite{5}. Using pulsed laser depositions (PLD), Ueda {\it et al.}
reported ferromagnetism (FM) above room temperature\cite{6}, while Jin {\it %
et al.} found no indication of FM by utilizing laser molecular beam epitaxy%
\cite{7}. This controversy between research teams may result from the growth
method used and/or from the growth conditions (oxygen pressure, deposition
temperature, etc...). In the particular case of the PLD technique, it may
also arise from the targets preparation and this parameter has never been
considered up to now. One of the reason is that the control of the dopant
incorporation would be quite difficult to obtain using a pre-doped ceramic
oxide target\cite{10}. This is a crucial point since the properties of the
DMS are very sensitive to the percentage of dopant\cite{11}. The homogenity
of dopant incorporation as well as the precise control of the growth might
be responsible for the changes in the physical properties of the films
obtained by the different groups.

Therefore, the objective of this investigation is two fold : first to
develop an accurate method to grow the Co:ZnO films with a precise doping
and second to understand their properties. To achieve such a goal, Co-doped
ZnO\ films were deposited from two pure metal targets of Zn and Co and our
results are reported in this letter.

The Co:ZnO films were grown using the pulsed laser deposition technique.
Zinc (99.995\%) and Cobalt (99.995\%) targets were used as purchased (NEYCO,
France) without further preparations. The films are deposited using a KrF
laser ( $\lambda =248mm$)\cite{12} on (0001) Al$_2$O$_3$ substrates. The
substrates were kept at a constant temperature in the range 500${{}^{\circ }}
$C-750${{}^{\circ }}$C during the deposition which was carried out a
pressure around $0.1Torr$ of pure oxygen. After deposition, the samples were
slowly cooled to room temperature at a pressure of $300mTorr$ of O$_2$. The
deposition rate is $3Hz$ and the energy density is close to $2J/cm^2$. The
composition of the film was checked and corresponds to the nominal one in
the limit of the accuracy.

The structural study was done by X-Ray diffraction (XRD) using a seifert XRD
3000P for the $\Theta -2\Theta $ scans and an X'Pert Phillips for the
in-plane measurements (Cu, K$\alpha 1$, $\lambda =0.15406nm$).

In order to grow Co:ZnO films, we firstly need to deposit high quality ZnO
films. The resulting XRD pattern of ZnO is show in Fig.1. The two
diffractions peaks observed around $34.48{{}^{\circ }}$ and $72.66{{}^{\circ
}}$ are characteristic of the hexagonal ZnO wurzite, the $c$-axis being
perpendicular to the substrate plan. The out-of-plane lattice parameter is
calculated to be 0.52mm which corresponds to the theoretical bulk one \cite
{13}. The sharp and intense peaks observed indicate that the films are
highly crystallized which is confirmed by the low value of the full-width at
half maximum (FWHM) of the rocking curve recorded around the $002$ reflexion
(see inset of Fig1). The epitaxial relationships between ZnO films and Al$_2$%
O$_3$ substrates are determined using asymmetrical XRD. Fig.2 displays the $%
\Phi -$scan of the ZnO films obtain from the ($103$) planes. The peaks are
separated by 60${{}^{\circ }}$, indicating a six-fold symmetry with a
rotation of 30${{}^{\circ }}$ of the ZnO symmetry in the plane, with respect
to the sapphire substrate. The averaged in-plane lattice parameter of ZnO is
0.325mm and the FWHM of the peaks in the $\Phi -$scan of ZnO is small ($0.68{%
{}^{\circ }}$). This value is close to previous value reported in the
literature\cite{14}. In order to obtain additionnal information, on the
structural properties of the ZnO films, we have determined the thin film
strains. This technique used the distance between atomic plane of a
crystalline specimen as an internal strain gage\cite{15}. The plane spacing $%
d_{hkl}$ is normal to the diffraction vector $\overrightarrow{L}$. One can
define a strain $\varepsilon $ , along this diffraction vector $d_{hkl}$, $%
\varepsilon =(d_{hkl}-d_0)/d_0$ where $d_0$ is the unstressed plane spacing
of the ($hkl$) planes (i.e the value of the bulk). To measure the stress, we
used the sin$^2\Psi $ technique\cite{15}. Briefly, in this model, the strain
is defined as follows: $\varepsilon =\alpha \sin ^2\Psi $+$\beta $ (where $%
\alpha $ and $\beta $ are constantes that depends on the strain along the
the surface direction, the Young's modulus, Poisson's ratio and the stress
along the direction, for details see Ref.15). Fig.3 shows the evolution of
the strain $\varepsilon $\ as a function of $sin^2\Psi $ for ZnO\ films
grown at different temperatures. The strain increases with the temperature,
indicating that the film is more strained along the in-plane direction
(since the out-of-plane lattice parameter is practically constant whatever
the growth conditions). A similar conclusion is obtained when\ $\varepsilon $
is plot for various oxygen pressure (not shown). To minimize the
substrate-induced strain of the ZnO, we deposited the film at 600${{}^{\circ
}}C$ under $0.1Torr$ of O$_2$ (under these conditions $\varepsilon $\ is
almost constant as a function of $sin^2\Psi $).

These conditions will be used hereafter to synthesise Co-doped\ ZnO films.
We used the following procedure. For example, in order to grow a Co-doped
ZnO film, we fired $m$ pulse on Co and $n$ pulses on Zn. The sequence is
repeated until the desired thickness is obtained. Various compositions have
been grown with different $m/n$ ratios given the same results and indicating
a good reproducibility of the films. Moreover, in each case the film is a
single phase, highly crystallized (the FWHM\ is always around $0.25{%
{}^{\circ }}$). In the XRD patterns, only the diffraction peaks
corresponding to the Co:ZnO phase are observed suggesting that the Co
clusters are not present. This is confirmed by the x-rays topography
recorded in a scanning transmission electron microscopy (JEOL 2010F) of the
Co:ZnO films where an homogenous repartition of the Co is observed. The
details of the structural and microstructural characterizations will be
present elsewhere. Moreoever, it has been shown that the high pressure of
oxygen ($0.1Torr$ in the present case) reduces the Co clusters formation\cite
{16}. The out-of-plane lattice parameter of the Co:ZnO\ films increases
almost lineraly as a function of the Co doping and nearly obeys the
Vergard's law (Fig.4). Reagrding this curve, it seems that the limit of the
solution is close to 9\%, since for higher Co content, the lattice parameter
do not change.

We investigated the magnetic properties of these thin film samples using a
SQUID\ magnetometer. Fig.5 shows the $M(T)$ recorded for a 5\% Co-doped ZnO
film\cite{16a}. The ferromagnetic behavior is observed on the $M(H)$ in the
whole temperature range between 5-300K (inset of Fig.5). The hysteresis of
the magnetization is very small (about few Gauss). $M(T)$ curves (Fig.5)
clearly evidences that the film is ferromagnetic with a Curie temperature
around 300K. The transition from the ferromagnetic state to the paramagnetic
state is clearly seen, suggesting that the mettalic Co clusters (the $T_C$\
of the metal Co clusters is above 1000K) are not responsible for the effect
observed at 300K \cite{8,9}.\ Moreover, the saturation moment (0.7 $\mu _B$%
/mole Co) is very weak compared to $1.7\mu _B$ of metallic Co$^{[0]}$,
suggesting that the Co state should be close to Co$^{2+}$.\ The increase of
the out-of-plane lattice parameter as the cobalt content increases is also
in favor of Co$^{2+}$ \cite{17}. We believe that this is due to the
technique used in the study where not only the conditions of the deposition
minimize the strains but also the alternately deposition from the two
targets favors the homogeneity of the doped films. Moreover, it has been
seen that the low temperature of deposition leads to homogenous films\cite{8}%
.

In conclusion, we have developped an alternative method for the growth of
pulsed laser deposited thin films. This method permits an accurate control
of the dopant in the matrix. To illustrate this procedure, we firstly
synthesized high quality ZnO thin films on Al$_2$O$_3$ (0001) substrates and
minimized the substrate-induced strain by optimizing the growth conditions.
Secondly, we utilized this process to deposit high quality Co-doped ZnO\
films with a Curie temperature close to room temperature. The growth of
these ferromagnetic films opens the route for the fabrication of spin-based
electronics since this original method can be used to grow various oxide
thin films.

We acknowledge Nicole Prellier in the preparation of the manuscript.

Figures Captions:

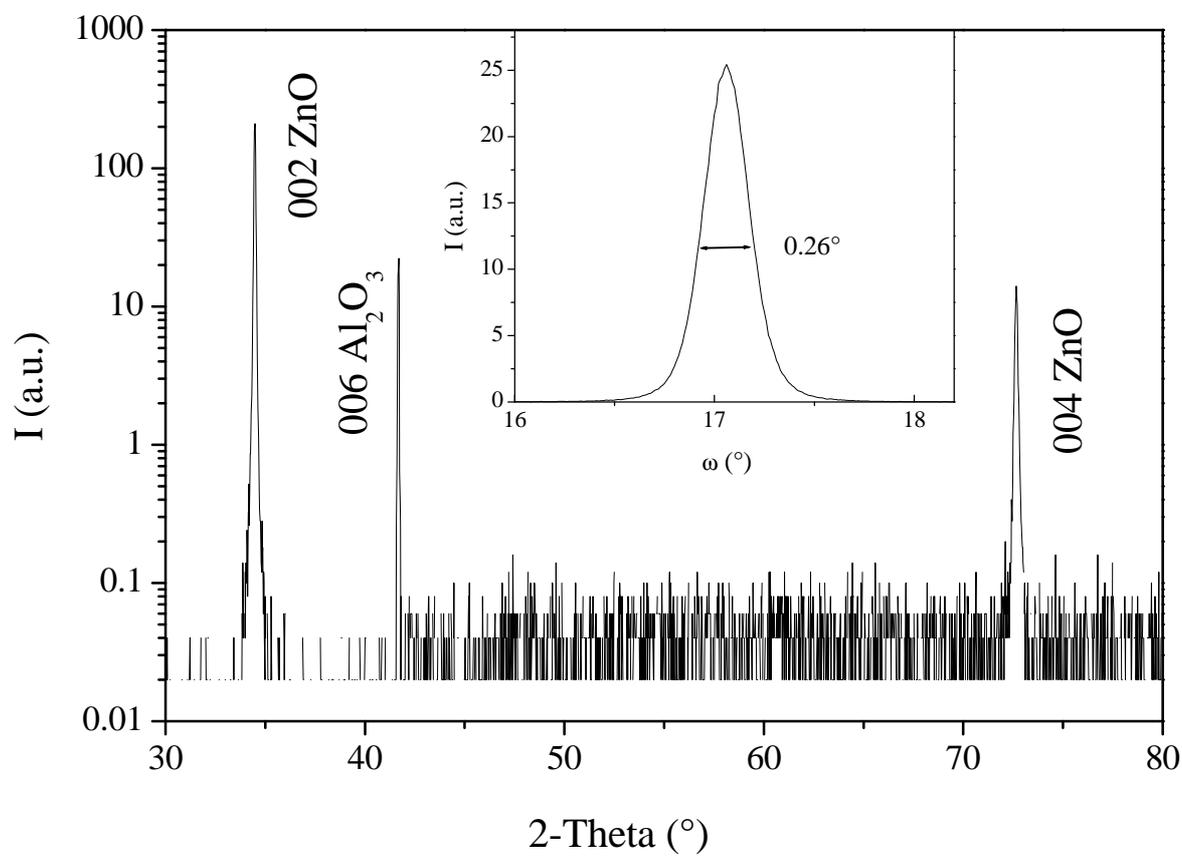
Figure 1: Room temperature $\Theta -2\Theta $ XRD pattern of typical ZnO
film. The inset depicts the rocking curve ($\omega $-scan) of (002)
reflection of the film. Note the sharpness and the high intensity of the
peaks.

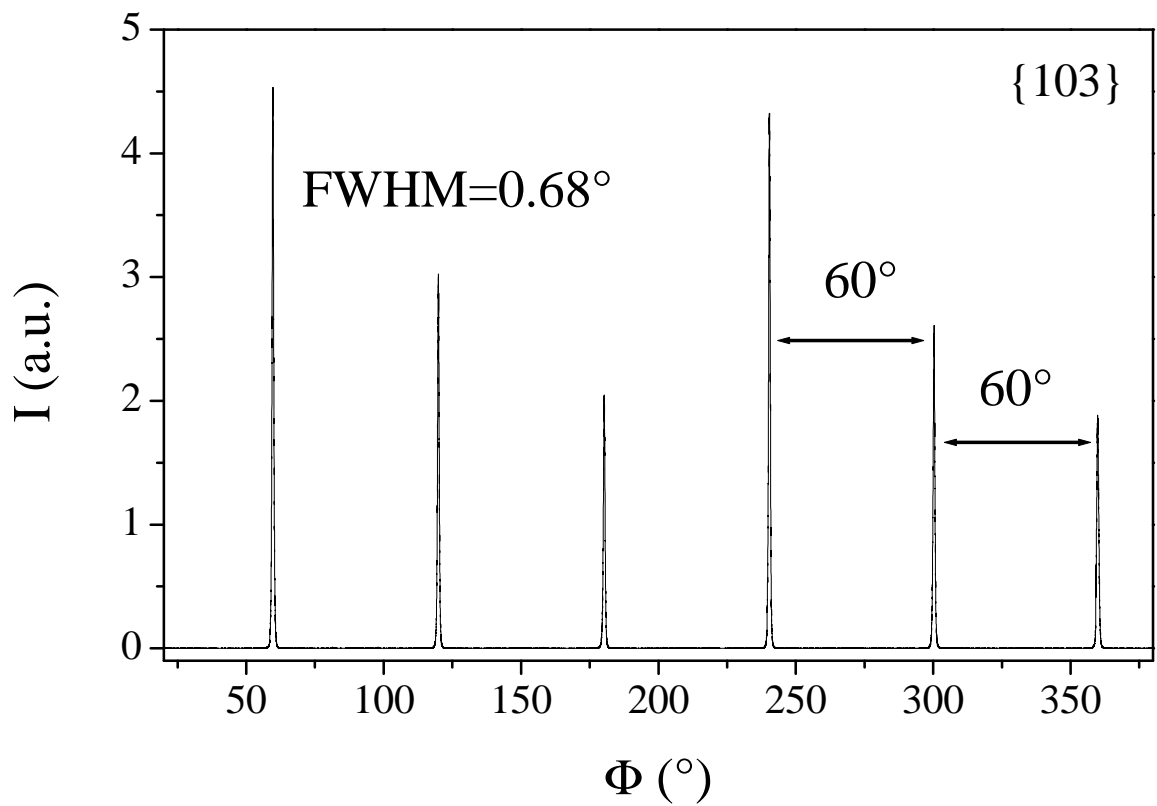
Figure 2: $\Phi $-scan of the \{103\} family peaks of a typical ZnO film
showing a very good in-plane orientation of the film.

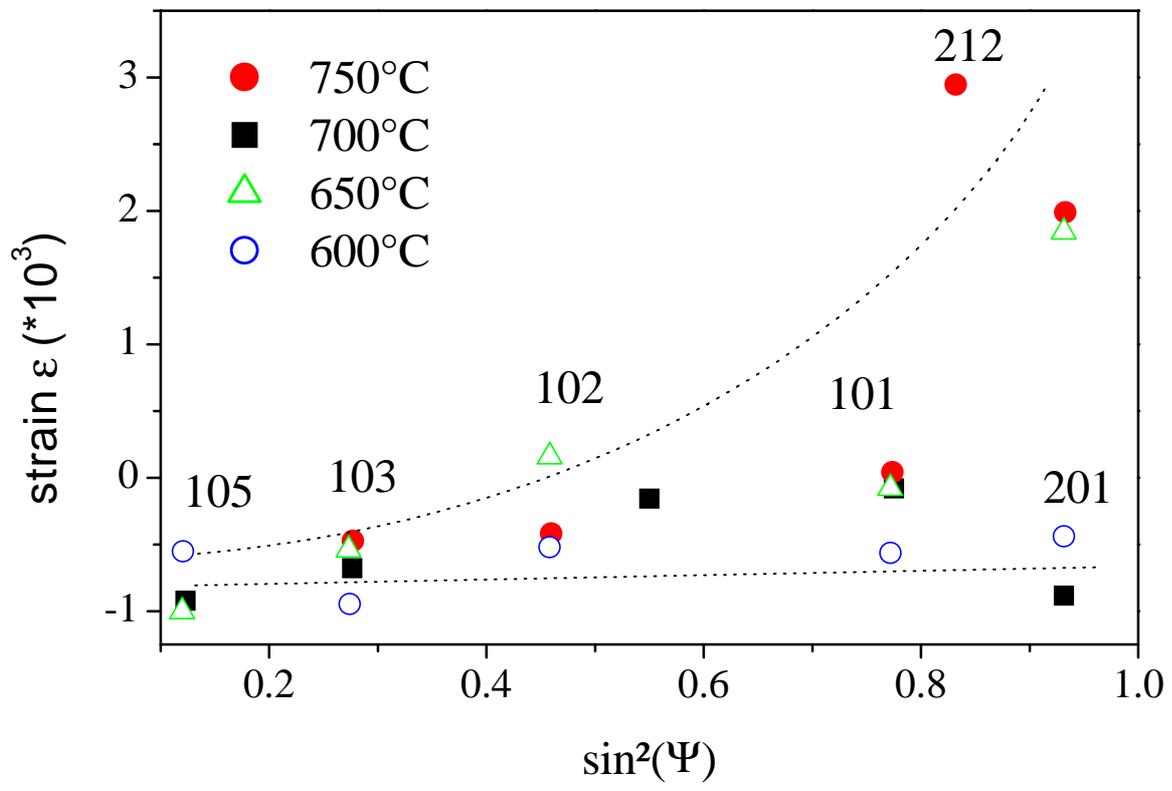
Figure 3: $\varepsilon $ vs. $sin^2\Psi $ for a ZnO\ films grown at
different deposition temperatures. The values of $(hkl)$ planes
corresponding to the measured $sin^2\Psi $ is indicating. The dot lines are
only a guide for the eyes.

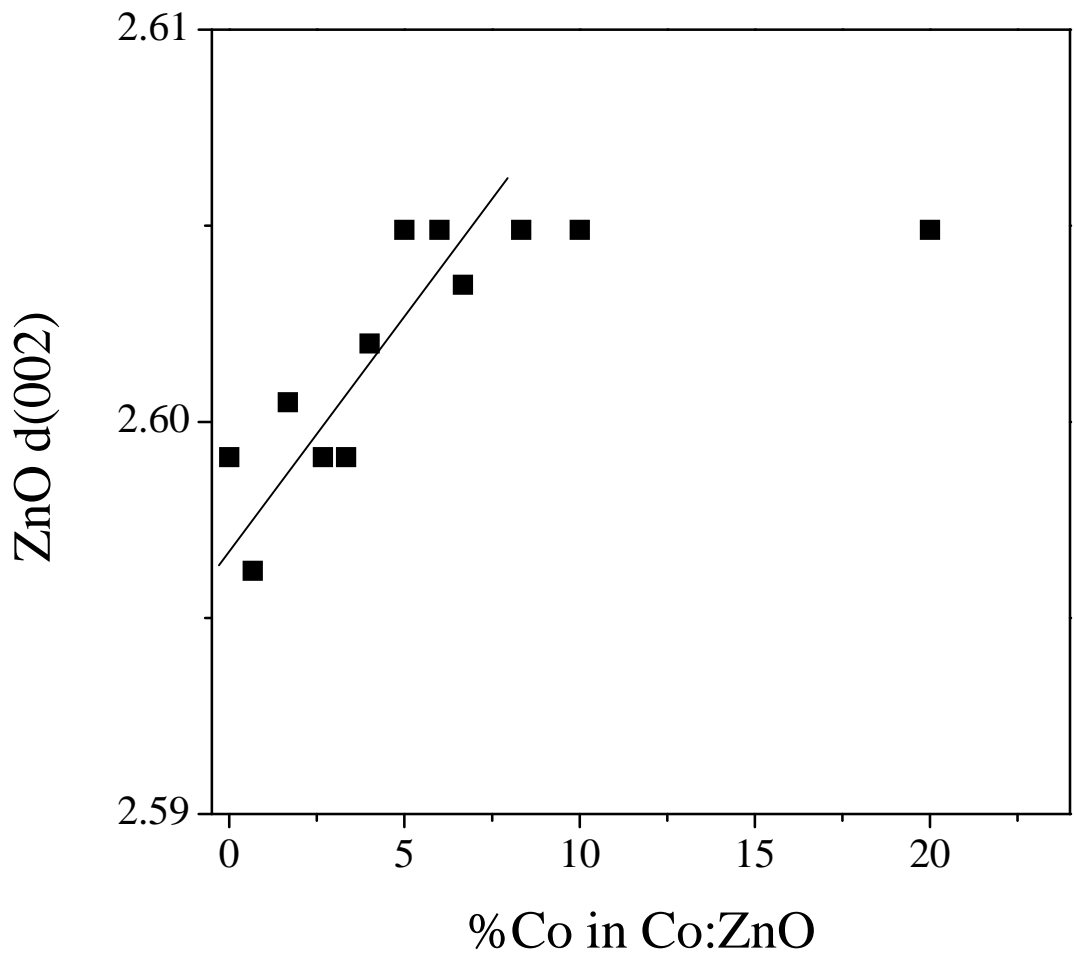
Figure 4: Evolution of the $d_{002}$ in Co:ZnO films as a function of the Co
content. The line is a guide for the eyes. The plateau above 15\% indicates
that thee limit of the solid solution is around 9\%.

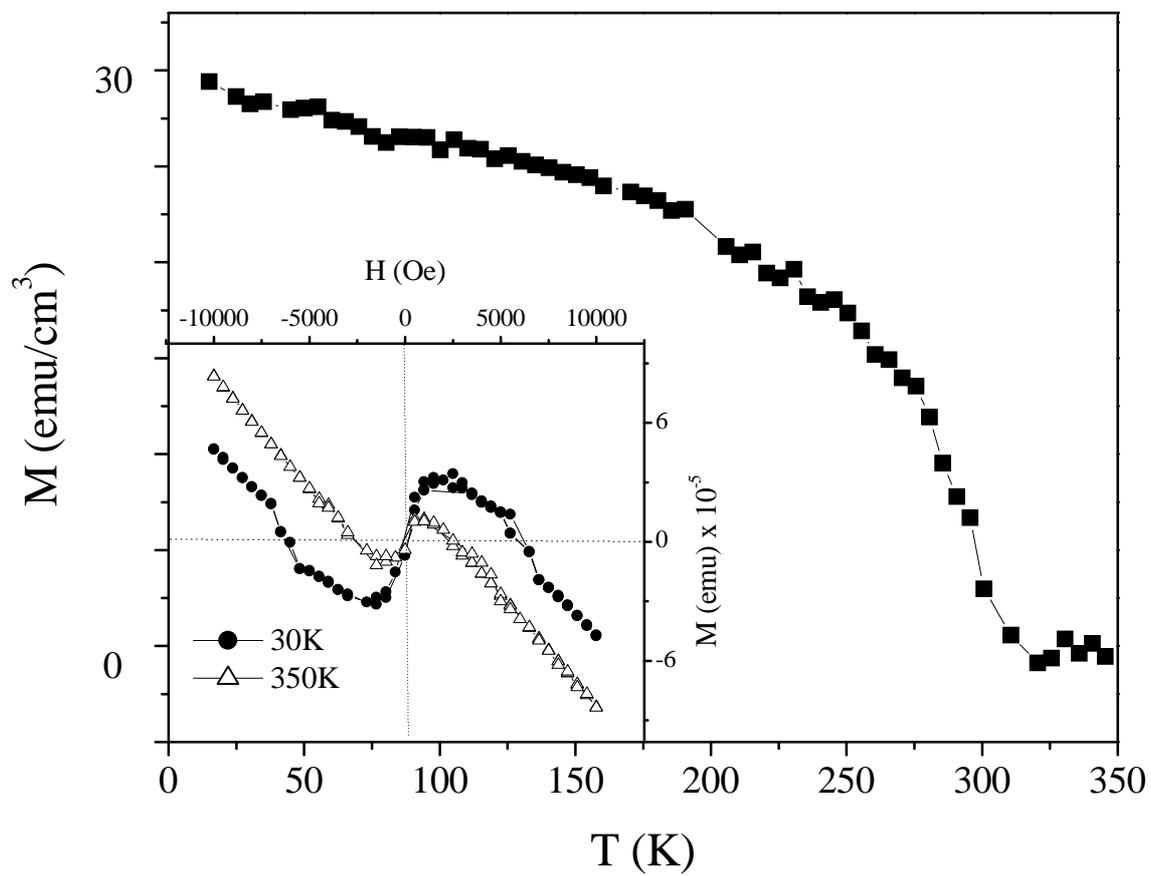
Figure 5: $M(T)$ of the Co:ZnO films (5\% of Co) recorded under $2000G$. The
inset depicts the $M(H)$ at 30K and 350K. The 350K signal is due to the
background component of the substrate. The anomaly at 30K $\pm 7000G$
corresponds to the difficulty in fitting the signal when crosses zero.


\begin{references}
\bibitem{16a}  We try to investigate the magnetic properties of samples with
a lower content of Co (%
\mbox{$<$}%
5\%). But in this case, the signal of the substrate is higher than the film
signal which makes impossible to conclude about a magnetic state of 2\% or
4\% Co-doped ZnO films.
\end{references}
\end{document}